\RequirePackage{fix-cm}

\documentclass[smallextended]{svjour3}       % onecolumn (second format)
\usepackage{amsfonts,amssymb,amsmath}
\usepackage{graphics,graphicx,epsfig}
\usepackage[dvipsnames]{xcolor}
\usepackage[normalem]{ulem}
\usepackage{comment}
\usepackage{booktabs}
\usepackage{epstopdf}
\usepackage{graphicx}
\usepackage{subfigure}
\usepackage{marvosym}
\usepackage{hyperref}
\usepackage{geometry}
\usepackage{cite}
\usepackage [latin1]{inputenc}

\def\be{\begin{eqnarray}}
\def\ee{\end{eqnarray}}
\def\ba{\begin{array}{l}}
\def\ea{\end{array}}

\begin{document}
\title{Detection of entanglement for multipartite quantum states}
\author{Hui Zhao$^1$ \and Yu-Qiu Liu $^1$ \and Naihuan Jing$^{2}$ \and Zhi-Xi Wang$^3$}
\institute{ \Letter~~Hui Zhao \at
                     zhaohui@bjut.edu.cn \and
              \Letter~~Yu-Qiu Liu \at
                     1091960127@qq.com \and
                           \Letter~~    Naihuan Jing\at
                    ~~~ jing@math.ncsu.edu \and
                     \Letter~~     Zhi-Xi Wang\at
                     ~~~wangzhx@cnu.edu.cn \and
           \and
\at 1 Department of Mathematics, Faculty of Science, Beijing University of Technology, Beijing 100124, China\\
\at 2 Department of Mathematics, North Carolina State University, Raleigh, NC 27695, USA\\
\at 3 School of Mathematical Sciences,  Capital Normal University,  Beijing 100048,  China
}

\maketitle
\begin{abstract}
\baselineskip18pt
We study genuine tripartite entanglement and multipartite entanglement of arbitrary $n$-partite quantum states by using the representations with generalized Pauli operators of a density matrices. While the usual Bloch representation of a density matrix uses three types of generators in the special unitary Lie algebra $\mathfrak{su}(d)$, the representation with generalized Pauli operators has one uniformed type of generators and it simplifies computation. In this paper, we take the advantage of this simplicity to derive useful and operational criteria to detect genuine tripartite
entanglement. We also obtain a sufficient criterion to detect entanglement for multipartite quantum states in arbitrary dimensions. The new method can detect more  entangled states than previous methods as backed by detailed examples.

\end{abstract}

\section{Introduction}
Quantum entanglement is a key resource in quantum information with wide applications in entanglement swapping \cite{bvk}, quantum cryptography \cite{eak} and quantum secure communication \cite{bw}. The genuine multipartite entanglement (GME) stands out
with significant properties \cite{hp,tg}. Thus measuring and detection of genuine multipartite entanglement for given states has been an important task in quantum computation.

A lot of methods have been presented in detecting
entanglement and genuine entanglement \cite{pa,hhh,hong,skt}. For tripartite quantum states, sufficient conditions to test entanglement of three-qubit states in the vicinity of the GHZ, the W states and the PPT entangled states were found in  \cite{akb}. A sufficient criterion for the entanglement of tripartite systems based on local sum uncertainty relations was proposed in \cite{ymm}. The sufficient conditions for judging genuine tripartite entanglement were presented by using partial transposition and realignment of density matrices in \cite{mes}. Yang et al \cite{ysc} derived a criterion for detecting genuine tripartite entanglement based on quantum Fisher information. By using the Bloch representation of density matrices and the norms of correlation tensors, the genuine tripartite entangled criteria were presented in \cite{lmj,dgh}. The authors in \cite{zzj} studied the separability criteria in tripartite and four-partite quantum system by the matrix method. The separability criteria for four-partite quantum system based on the upper bound of Bloch vectors were discussed in \cite{lww}. For higher dimensional quantum system, Chen et al \cite{cw} presented a generalized partial separability criterion of multipartite quantum systems in arbitrary dimensions. The separable criteria and $k$-separable criteria for general $n$-partite quantum states were given in \cite{hgy,lwf,xzz}.

Many of these methods have used the Bloch representation of the density matrix, which has
become more complex as dimension of the quantum system increases. This is partly due to the
fact that the Bloch representation is relied on the Gell-Mann basis of the special unitary Lie algebra $\mathfrak{su}(d)$
which has three kinds of basis elements: upper, diagonal and lower matrices.  In view of this, perhaps
using another well-known basis of the Lie algebra $\mathfrak{su}(d)$: the Weyl basis to study quantum entanglement will likely simplify some of the criteria, as the latter consists of uniformed basis elements.  In Ref. \cite{bgj}, the authors showed that the principal basis matrix plays an essential role in the representation theory of the Yangian $Y(\mathfrak{sl}(3))$ which has a close relation with the study of entangled states in quantum information (see also for recent applications \cite{fss}).

In this paper, we study entanglement of multipartite quantum systems by using the representation with generalized Pauli operators, and we obtain several better criteria in detecting the GME than previously available tests. The paper is organized as follows. In section 2, after reviewing the representation with generalized Pauli operators of the quantum state, we construct matrices by using the correlation tensors and derive the criteria to detect entanglement and genuine tripartite entanglement. By detailed example, our results can detect more genuine entangled states. In section 3, we obtain the entanglement theorem for arbitrary $n$-partite quantum systems. Conclusions
are given in section 4.

\section{Genuine entanglement for tripartite quantum state}
We first consider the GME for tripartite states. Let $E_{ij}$ be the $d\times d$ unit matrix with the only nonzero entry 1 at the position $(i,j)$, and let $\omega$ be a fixed $d$-th primitive root of unity. By means of the division with remainder, for certain $d_s$ and $u_s\in\{0,\cdots,d_s^2-1\}$, there exists unique integers $i$ and $j$ such that $u_s=d_si+j$ $(0\leq i,j\leq d_s-1)$, then the generalized Pauli operators of the $s$th $d_s$-dimensional Hilbert space $H_s^{d_s}$ are given by
\begin{equation}\label{1}
A_{u_s}^{(s)}=A_{d_si+j}^{(s)}=\sum\limits_{m=0}^{d_s-1}\omega^{im}E_{m,m+j},
\end{equation}
where $\omega^{d_s}=1$. The basis obeys the algebraic relation:
\begin{equation*}
  A_{d_si+j}^{(s)}A_{d_sk+l}^{(s)}=\omega^{jk}A_{d_s(i+k)+(j+l)}^{(s)},
\end{equation*}
then $(A_{d_si+j}^{(s)})^{\dagger}=\omega^{ij}A_{d_s(d_s-i)+(d_s-j)}^{(s)}$, so $tr(A_{d_si+j}^{(s)}(A_{d_sk+l}^{(s)})^{\dagger})=\delta_{ik}\delta_{jl}d_s$\cite{hjz}, where $\dagger$ stands for conjugate transpose.
Denote by $\|\cdot\|$ the norm of a (column) complex vector, i.e. $\|v\|=\sqrt{v^{\dagger} v}$.
The trace norm (Ky Fan norm) of a retangular matrix $A\in \mathbb{C}^{m\times n}$ is defined as $\|A\|_{tr}=\sum\sigma_i=tr\sqrt{AA^\dagger}$,
where $\sigma_i$ are the singular values of $A$ and $\|A\|_{tr}\leq\sqrt{\mathrm{min}\{m,n\}}\|A\|$ for any matrix $A$. Clearly $\|A\|_{tr}=\|A^{\dagger}\|_{tr}$.

\begin{lemma}\label{lemma:1} Let $H_s^{d_s}$ denote the $s^{th}$ $d_s$-dimensional Hilbert space. For a quantum state $\rho_1\in H_1^{d_1}$, $\rho_1$ can be expressed as
$\rho_1=\frac{1}{d_1}\sum\limits_{u_1=0}^{d_1^2-1}t_{u_1}A_{u_1}^{(1)}$,
where $A_0^{(1)}=I_{d_1}$, $t_{u_1}=tr(\rho_1(A_{u_1}^{(1)})^{\dagger})$ are complex coefficients. Let $T^{(1)}$ be the column vector with entries $t_{u_1}$ for $u_1\neq0$,
we have
\begin{equation}\label{2}
\|T^{(1)}\|^2\leq d_1-1.
\end{equation}
\end{lemma}
{\it Proof}~~Since $tr(\rho_1^2)\leq1$, we have
$$tr(\rho_1^2)=tr(\rho_1\rho_1^{\dagger})=\frac{1}{d_1}(1+\|T^{(1)}\|^2)\leq1,$$
namely, $\|T^{(1)}\|^2\leq d_1-1$.
\qed

For a state $\rho_{12}\in H_1^{d_1}\otimes H_2^{d_2}$,
$\rho_{12}$ has the generalized Pauli operators representation:
\begin{equation}\label{3}
\rho_{12}=\frac{1}{d_1d_2}\sum\limits_{u_1=0}^{d_1^2-1}
\sum\limits_{u_2=0}^{d_2^2-1}t_{u_1,u_2}A_{u_1}^{(1)}\otimes A_{u_2}^{(2)}
\end{equation}
where $A_0^{(s)}=I_{d_s}(s=1,2)$, the coefficients
$t_{u_1,u_2}=tr(\rho_{12}(A_{u_1}^{(1)})^{\dagger}\otimes (A_{u_2}^{(2)})^{\dagger})$ are complex numbers.
Let $T^{(1)}$, $T^{(2)}$, $T^{(12)}$ be the vectors with entries $t_{u_1,0}$, $t_{0,u_2}$, $t_{u_1,u_2}$ for $u_1,u_2\neq0$.

\begin{lemma}\label{lemma:2}
Let $\rho_{12}\in H_1^{d_1}\otimes H_2^{d_2}$ be a mixed state, we have $\|T^{(12)}\|^2\leq d_1d_2(1-\frac{1}{d_1^2}-\frac{1}{d_2^2})+1$.
\end{lemma}
{\it Proof}~~For a pure state $\rho_{12}$, we have $tr(\rho_{12}^2)=1$, namely
\begin{equation}\label{4}
tr(\rho_{12}^2)=tr(\rho_{12}\rho_{12}^{\dagger})
=\frac{1}{d_1d_2}(1+\|T^{(1)}\|^2+\|T^{(2)}\|^2+\|T^{(12)}\|^2)
=1.
\end{equation}
By using $tr(\rho_1^2)=tr(\rho_2^2)$, we have $\frac{1}{d_1}(1+\|T^{(1)}\|^2)=\frac{1}{d_2}(1+\|T^{(2)}\|^2)$, where $\rho_1$ and $\rho_2$ are the reduced density operators on $H_1^{d_1}$ and $H_2^{d_2}$, respectively. Then
\begin{equation*}
\frac{1}{d_1^2}(1+\|T^{(1)}\|^2)+\frac{1}{d_2^2}(1+\|T^{(2)}\|^2)=
\frac{1}{d_1d_2}(2+\|T^{(1)}\|^2+\|T^{(2)}\|^2).
\end{equation*}
Using the above two equations we obtain that
\begin{equation*}
\begin{split}
\|T^{(12)}\|^2&=d_1d_2-1-\|T^{(1)}\|^2-\|T^{(2)}\|^2\\
&=d_1d_2+1-[\frac{d_2}{d_1}(1+\|T^{(1)}\|^2)+\frac{d_1}{d_2}(1+\|T^{(2)}\|^2)].
\end{split}
\end{equation*}
By $\|T^{(1)}\|^2\geq0$, $\|T^{(2)}\|^2\geq0$, we have
\begin{equation}\label{5}
\begin{split}
\|T^{(12)}\|^2&\leq d_1d_2+1-(\frac{d_2}{d_1}\cdot1+\frac{d_1}{d_2}\cdot1)\\
&=d_1d_2(1-\frac{1}{d_1^2}-\frac{1}{d_2^2})+1,
\end{split}
\end{equation}
If $\rho$ is a mixed state, then $\rho=\sum_{i}p_i\rho_i$ is a convex sum of pure states, $\sum_ip_i=1$.
Then $\|T^{(12)}(\rho)\|\leq \sum_ip_i\|T^{(12)}(\rho_i)\|\leq \sqrt{d_1d_2(1-\frac{1}{d_1^2}-\frac{1}{d_2^2})+1}.$ \qed

A general tripartite state $\rho\in H_{1}^{d_1}\otimes H_{2}^{d_2}\otimes H_{3}^{d_3}$ can be written in terms of the generalized Pauli operators:
\begin{equation}\label{6}
\rho=\frac{1}{d_1d_2d_3}\sum_{u_1=0}^{d_1^2-1}\sum_{u_2=0}^{d_2^2-1}\sum_{u_3=0}^{d_3^2-1}t_{u_1,u_2,u_3}A_{u_1}^{(1)}\otimes A_{u_2}^{(2)}\otimes A_{u_3}^{(3)},
\end{equation}
where $A_{u_f}^{(f)}$ stands for that the tensor operator with $A_{u_f}$ acting on the space $H_f^{d_f}$, $A_{0}^{(f)}=I_{d_f}$, $t_{u_1,u_2,u_3}=tr(\rho (A_{u_1}^{(1)})^{\dagger}\otimes (A_{u_2}^{(2)})^{\dagger}\otimes (A_{u_3}^{(3)})^{\dagger})$ are the
complex coefficients.
%Let $T^{(f)}$ and $T^{(fg)}$ be the (column) vectors with entries of $t_{u_f,0,0}$ and $t_{u_f,u_g,0}$ for $u_f, u_g\neq0$, $f,g\in\{1,2,3\}$.

In the following we will construct some matrices out of the expansion coefficients of the density matrix $\rho$ in \eqref{6}. For $f,g,h\in\{1,2,3\}$, set
\begin{equation}\label{7}
N^{f|gh}=diag\{N^{f|gh}_1, N^{f|gh}_2, \cdots, N^{f|gh}_{d_f^2-1}\},
\end{equation}
where $N^{f|gh}_i=[t_{i,u_g,u_h}]$ is a ${(d_g^2-1)\times (d_h^2-1)}$ matrix, $i=1,2,\cdots, d_f^2-1$. For example, when $\rho\in H_1^2\otimes H_2^2\otimes H_3^3$, $N^{2|13}=diag\{N^{2|13}_1, N^{2|13}_2, N^{2|13}_3\}$, where
$$N_i^{2|13}=\left[
  \begin{array}{cccc}
    t_{1,i,1} & t_{1,i,2} & \cdots & t_{1,i,8} \\
    t_{2,i,1} & t_{2,i,2} & \cdots & t_{2,i,8} \\
    t_{3,i,1} & t_{3,i,2} & \cdots & t_{3,i,8} \\
  \end{array}
\right], i=1,2,3.
$$

\begin{theorem}\label{1}
For a biseparable tripartite pure state $\rho\in H_1^{d_1}\otimes H_2^{d_2}\otimes H_3^{d_3}$ and  $f\neq g\neq h\in\{1,2,3\}$, we have\\
$(i)$~if $\rho$ is separable under bipartite partition $f|gh$ , then
$$\|N^{f|gh}\|_{tr}\leq\sqrt{(d_f^2-1)\cdot min\{d_g^2-1,d_h^2-1\}
(d_f-1)[d_gd_h(1-\frac{1}{d_g^2}-\frac{1}{d_h^2})+1]}$$
$(ii)$~if $\rho$ is separable under bipartite partition $g|fh$ , then
$$\|N^{f|gh}\|_{tr}\leq\sqrt{(d_f^2-1)
(d_g-1)[d_fd_h(1-\frac{1}{d_f^2}-\frac{1}{d_h^2})+1]}$$
$(iii)$~if $\rho$ is separable under bipartite partition $h|fg$ , then
$$\|N^{f|gh}\|_{tr}\leq\sqrt{(d_f^2-1)
(d_h-1)[d_fd_g(1-\frac{1}{d_f^2}-\frac{1}{d_g^2})+1]}$$
\end{theorem}
{\it Proof}~(i)~If the tripartite pure state $\rho$ is separable under the bipartition $f|gh$, then it can be expressed as
\begin{equation}\label{8}
\rho_{f|gh}=\rho_f\otimes \rho_{gh},
\end{equation}
where
\begin{equation}\label{9}
\rho_f=\frac{1}{d_f}\sum\limits_{u_f=0}^{d_f^2-1}t_{u_f}A_{u_f}^{(f)},
\end{equation}
\begin{equation}\label{10}
\begin{split}
\rho_{gh}=\frac{1}{d_gd_h}\sum\limits_{u_g=0}^{d_g^2-1}\sum\limits_{u_h=0}^{d_h^2-1}t_{u_g,u_h}A_{u_g}^{(g)}\otimes A_{u_h}^{(h)}.
\end{split}
\end{equation}
Let $T^{(f)}$, $T^{(g)}$, $T^{(h)}$ and $T^{(gh)}$ be the vectors with entries $t_{u_f}$, $t_{u_g,0}$, $t_{0,u_h}$ and $t_{u_g,u_h}$ for $u_f,u_g,u_h\neq0$.
Then
\begin{equation}\label{13}
N^{f|gh}=\left[
             \begin{array}{cccc}
               t_1 & \  & \  & \  \\
               \  & t_2 & \  & \  \\
               \  & \  & \ddots & \  \\
               \  & \  & \  & t_{d_f^2-1} \\
             \end{array}
           \right]\otimes\left[
             \begin{array}{cccc}
               t_{11}   & t_{12}    & \cdots & t_{1(d_h^2-1)} \\
               t_{21}   & t_{22}    & \cdots & t_{2(d_h^2-1)} \\
               \vdots & \vdots  & \      & \vdots \\
               t_{(d_g^2-1)1} & t_{(d_g^2-1)2}  & \cdots & t_{(d_g^2-1)(d_h^2-1)} \\
             \end{array}
           \right].
\end{equation}
%where $\dagger$ stands for conjugate transpose.
It follows from Lemma 1 and Lemma 2 that
\begin{equation*}
\begin{split}
\|N^{f|gh}\|_{tr}&=\|\left[
             \begin{array}{cccc}
               t_1 & \  & \  & \  \\
               \  & t_2 & \  & \  \\
               \  & \  & \ddots & \  \\
               \  & \  & \  & t_{d_f^2-1} \\
             \end{array}
             \right]\|_{tr}
  \|\left[
             \begin{array}{ccc}
               t_{11}    & \cdots & t_{1(d_h^2-1)} \\
               \vdots    & \      & \vdots \\
               t_{(d_g^2-1)1} & \cdots & t_{(d_g^2-1)(d_h^2-1)} \\
             \end{array}
           \right]\|_{tr}\\
&\leq \sqrt{(d_f^2-1)}\|T^{(f)}\|\cdot \sqrt{min\{d_g^2-1,d_h^2-1\}}\|T^{(gh)}\|\\
&\leq \sqrt{(d_f^2-1)\cdot min\{d_g^2-1,d_h^2-1\}
(d_f-1)[d_gd_h(1-\frac{1}{d_g^2}-\frac{1}{d_h^2})+1]}.
\end{split}
\end{equation*}

(ii)~If the tripartite pure state $\rho$ is separable under the bipartition $g|fh$, then it can be expressed as
\begin{equation}\label{12}
\rho_{g|fh}=\rho_g\otimes \rho_{fh},
\end{equation}
where
\begin{equation}\label{13}
\rho_g=\frac{1}{d_g}\sum\limits_{u_g=0}^{d_g^2-1}t_{u_g}A_{u_g}^{(g)},
\end{equation}
\begin{equation}\label{14}
\begin{split}
\rho_{fh}=\frac{1}{d_fd_h}\sum\limits_{u_f=0}^{d_f^2-1}\sum\limits_{u_h=0}^{d_h^2-1}t_{u_f,u_h}A_{u_f}^{(f)}\otimes A_{u_h}^{(h)}.
\end{split}
\end{equation}
Then, we have $N_i^{f|gh}=T^{(g)}\cdot (T_i^{(fh)})^t$,
%~ i=1,2,\cdots d_f^2-1
and
\begin{equation*}
\begin{split}
\|N^{f|gh}\|_{tr}&=\|T^{(g)}\|\cdot \sum_{i=1}^{d_f^2-1}\|T_i^{(fh)}\|\leq\sqrt{d_f^2-1}\|T^{(g)}\|\|T^{(fh)}\|\\
&\leq \sqrt{(d_f^2-1)
(d_g-1)[d_fd_h(1-\frac{1}{d_f^2}-\frac{1}{d_h^2})+1]}
\end{split}
\end{equation*}
where $T_i^{(fh)}$ is the vector with entries $t_{i,u_h}, i=1,\cdots,d_f^2-1$, $t$ stands for transpose.

(iii)~Using similar method, if $\rho$ is separable under the bipartition $h|fg$, we have $N_i^{f|gh}=T_i^{(fg)}\cdot (T^{(h)})^t$, $i=1,\cdots,d_f^2-1$
and
$$\|N^{f|gh}\|_{tr}\leq\sqrt{(d_f^2-1)
(d_h-1)[d_fd_g(1-\frac{1}{d_f^2}-\frac{1}{d_g^2})+1]}.$$
\qed

%where we have used the triangular inequality of the norm $\| \cdot \|_{tr}$
%and $\||a\rangle\langle b|\|_{tr}=\||a\rangle\|\||b\rangle\|$
%for real vectors $|a\rangle$ and $|b\rangle$.

Now we consider genuine tripartite entanglement. A mixed state is said to be genuine multipartite entangled if it cannot be written as a convex combination of biseparable states. Let $T(\rho)=\frac{1}{3}(\|N^{1|23}\|_{tr}+\|N^{2|13}\|_{tr}+\|N^{3|12}\|_{tr})$, %after choosing certain (fixed) values of $\{\alpha, \beta\}$,
we define
\begin{equation*}
\begin{split}
Q_1=\textrm{Max}\{&\sqrt{(d_1^2-1)\cdot min\{d_2^2-1,d_3^2-1\}
(d_1-1)[d_2d_3(1-\frac{1}{d_2^2}-\frac{1}{d_3^2})+1]},\\ &\sqrt{(d_1^2-1)
(d_2-1)[d_1d_3(1-\frac{1}{d_1^2}-\frac{1}{d_3^2})+1]},\\ &\sqrt{(d_1^2-1)
(d_3-1)[d_1d_2(1-\frac{1}{d_1^2}-\frac{1}{d_2^2})+1]}\}.
\end{split}
\end{equation*}
%$$K_1=\textrm{Max}\{\sqrt{d_f-1}\left(|\alpha|\sqrt{d_g-1}+|\beta|\sqrt{d_gd_h(1-\frac{1}{d_g^2}-\frac{1}{d_h^2})+1}\right)\},$$
\begin{equation*}
\begin{split}
Q_2=\textrm{Max}\{&\sqrt{(d_2^2-1)\cdot min\{d_1^2-1,d_3^2-1\}
(d_2-1)[d_1d_3(1-\frac{1}{d_1^2}-\frac{1}{d_3^2})+1]},\\ &\sqrt{(d_2^2-1)
(d_1-1)[d_2d_3(1-\frac{1}{d_2^2}-\frac{1}{d_3^2})+1]},\\ &\sqrt{(d_2^2-1)
(d_3-1)[d_2d_1(1-\frac{1}{d_2^2}-\frac{1}{d_1^2})+1]}\}.
\end{split}
\end{equation*}
\begin{equation*}
\begin{split}
Q_3=\textrm{Max}\{&\sqrt{(d_3^2-1)\cdot min\{d_1^2-1,d_2^2-1\}
(d_3-1)[d_1d_2(1-\frac{1}{d_1^2}-\frac{1}{d_2^2})+1]},\\ &\sqrt{(d_3^2-1)
(d_1-1)[d_3d_2(1-\frac{1}{d_3^2}-\frac{1}{d_2^2})+1]},\\ &\sqrt{(d_3^2-1)
(d_2-1)[d_3d_1(1-\frac{1}{d_3^2}-\frac{1}{d_1^2})+1]}\}.
\end{split}
\end{equation*}
We have the following theorem.
\begin{theorem}\label{2}
A mixed state $\rho\in H_1^{d_1}\otimes H_2^{d_2}\otimes H_3^{d_3}$ is genuine tripartite entangled if $T(\rho)>\frac{1}{3}(Q_1+Q_2+Q_3)$.
\end{theorem}
{\it Proof}~ If $\rho$ is a biseparable, one has
$\rho=\sum_{i}o_{i}\rho_{i}^{1}\otimes\rho_{i}^{23}
+\sum_{j}r_{j}\rho_{j}^{2}\otimes\rho_{j}^{13}+\sum_{k}s_{k}\rho_{k}^{3}\otimes\rho_{k}^{12}$ with $0\leq o_{i},r_{j},s_{k}\leq1$ and $\sum_{i}o_{i}+\sum_{j}r_{j}+\sum_{k}s_{k}=1$. By Theorem 1, we have that
\begin{equation}
\begin{split}
&T(\rho)=\frac{1}{3}(\|N^{1|23}(\rho)\|_{tr}+\|N^{2|13}(\rho)\|_{tr}+\|N^{3|12}(\rho)\|_{tr})\\
=&\frac{1}{3}[\|N^{1|23}(\sum_{i}o_{i}\rho_{i}^{1}\otimes\rho_{i}^{23}
+\sum_{j}r_{j}\rho_{j}^{2}\otimes\rho_{j}^{13}+\sum_{k}s_{k}\rho_{k}^{3}\otimes\rho_{k}^{12})\|_{tr}+\|N^{2|13}(\sum_{i}o_{i}\rho_{i}^{1}\otimes\rho_{i}^{23}
\\+&\sum_{j}r_{j}\rho_{j}^{2}\otimes\rho_{j}^{13}+\sum_{k}s_{k}\rho_{k}^{3}\otimes\rho_{k}^{12})\|_{tr}+\|N^{3|12}(\sum_{i}o_{i}\rho_{i}^{1}\otimes\rho_{i}^{23}
+\sum_{j}r_{j}\rho_{j}^{2}\otimes\rho_{j}^{13}+\sum_{k}s_{k}\rho_{k}^{3}\otimes\rho_{k}^{12})\|_{tr}]\\
\leq&\frac{1}{3}[\sum_{i}o_{i}\|N^{1|23}(\rho_{i}^{1}\otimes\rho_{i}^{23})\|_{tr}+\sum_{j}r_{j}\|N^{1|23}(\rho_{j}^{2}\otimes\rho_{j}^{13})\|_{tr}+\sum_{k}s_{k}\|N^{1|23}(\rho_{k}^{3}\otimes\rho_{k}^{12})\|_{tr}\\
&+\sum_{i}o_{i}\|N^{2|13}(\rho_{i}^{1}\otimes\rho_{i}^{23})\|_{tr}+\sum_{j}r_{j}\|N^{2|13}(\rho_{j}^{2}\otimes\rho_{j}^{13})\|_{tr}+\sum_{k}s_{k}\|N^{2|13}(\rho_{k}^{3}\otimes\rho_{k}^{12})\|_{tr}\\
&+\sum_{i}o_{i}\|N^{3|12}(\rho_{i}^{1}\otimes\rho_{i}^{23})\|_{tr}+\sum_{j}r_{j}\|N^{3|12}(\rho_{j}^{2}\otimes\rho_{j}^{13})\|_{tr}+\sum_{k}s_{k}\|N^{3|12}(\rho_{k}^{3}\otimes\rho_{k}^{12})\|_{tr}]\\
\leq& \frac{1}{3}[(\sum_{i}o_{i}Q_1+\sum_{j}r_{j}Q_1+\sum_{k}s_{k}Q_1)+(\sum_{i}o_{i}Q_2+\sum_{j}r_{j}Q_2+\sum_{k}s_{k}Q_2)+(\sum_{i}o_{i}Q_3+\sum_{j}r_{j}Q_3+\sum_{k}s_{k}Q_3)\\
=&\frac{1}{3}[(\sum_{i}o_{i}+\sum_{j}r_{j}+\sum_{k}s_{k})Q_1+(\sum_{i}o_{i}+\sum_{j}r_{j}+\sum_{k}s_{k})Q_2+(\sum_{i}o_{i}+\sum_{j}r_{j}+\sum_{k}s_{k})Q_3\\
=&\frac{1}{3}(Q_1+Q_2+Q_3)\\
\end{split}
\end{equation}
Consequently, if $T(\rho)>\frac{1}{3}(Q_1+Q_2+Q_3)$, $\rho$ is genuine tripartite entangled.
\qed

Next we consider the permutational invariant state $\rho$, i.e. $\rho=\rho^p=p\rho p^{\dagger}$ for any permutation $p$ of the qudits. A biseparable permutational invariant state can be written as $\rho=\sum_i p_i\rho_i^1\otimes\rho_i^{23}+\sum_j r_j\rho_j^2\otimes\rho_j^{13}+\sum_k s_k\rho_k^3\otimes\rho_k^{12}$, where $0<p_i,r_j,s_k\leq1$. Set $d_1=d_2=d_3=d$, we have the following corollary.

\begin{corollary}
If a permutational invariant mixed state is biseparable, then we have
$$T(\rho)=\frac{1}{3}(\|N^{1|23}\|_{tr}+\|N^{2|13}\|_{tr}+\|N^{3|12}\|_{tr})\leq J_1.$$
Therefore if $T(\rho)>J_1$, $\rho$ is genuine tripartite entangled. Here
\begin{equation*}
J_1=\frac{(d-1)^2\sqrt{(d-1)(d^2-1)}+2(d^2-1)\sqrt{d-1}}{3}.
\end{equation*}
\end{corollary}

\begin{figure}[!htb]
  \centering
  \includegraphics[width=0.75\textwidth]{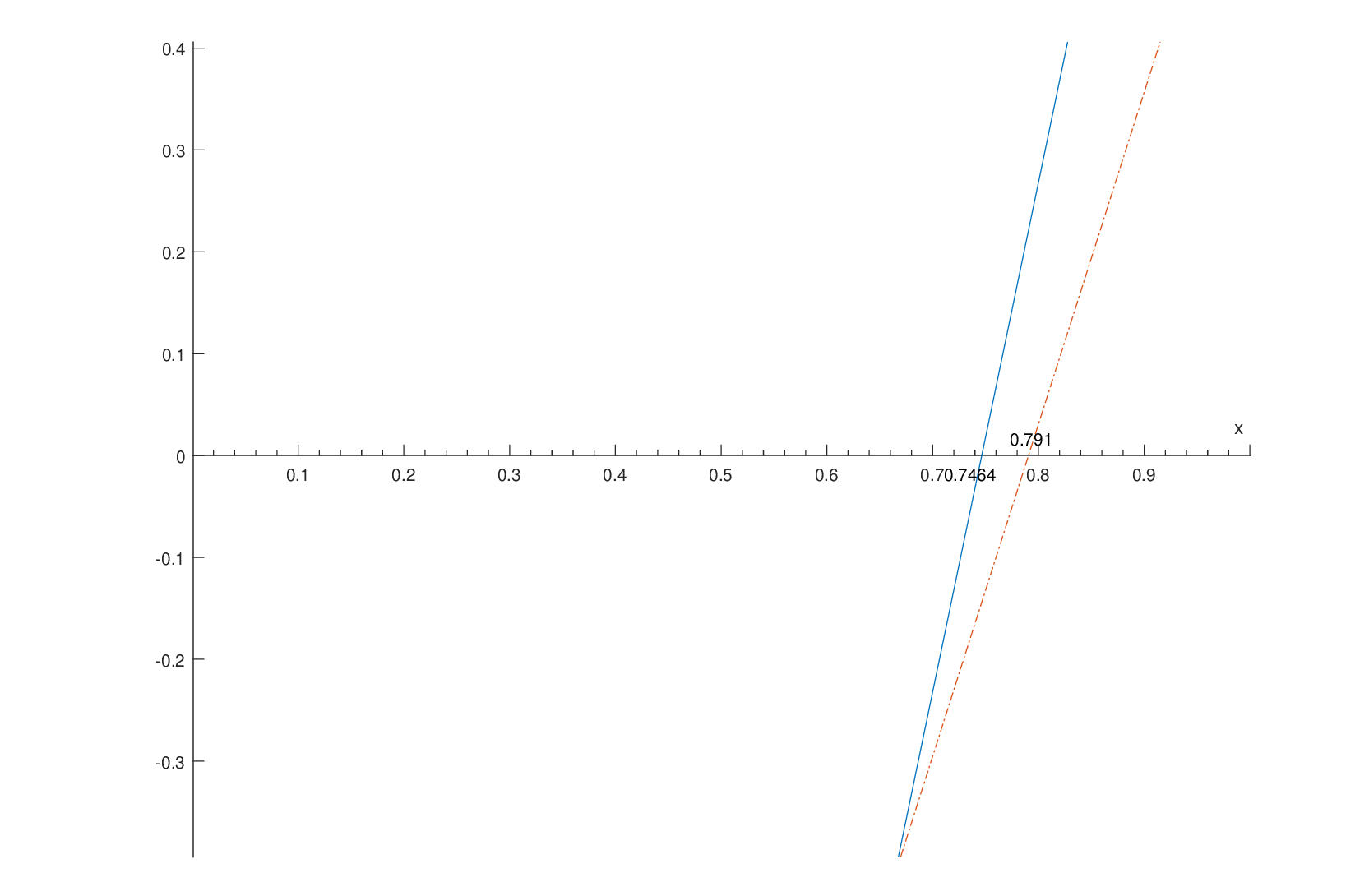}\\
  \caption{$f_1(x)$ from our result (solid straight line),
  %$g_1(x)$ from Theorem 2 in \cite{lmj}(dashed straight line),
  $g_1(x)$ from Theorem 2 in \cite{dgh}(dash-dot straight line). }
  \label{fig:1}
\end{figure}

\textit{\textbf{Example 1}} Consider the mixed three-qubit $W$ state,
\begin{equation}\label{14}
\rho=\frac{1-x}{8}I_8+x|W\rangle\langle W|, \quad 0\leq x\leq1,
\end{equation}
where $|W\rangle=\frac{1}{\sqrt{3}}(|001\rangle+|010\rangle+|100\rangle)$ and $I_8$ is the $8\times8$ identity matrix. %With different choices of $\alpha$ and $\beta$,
%Using Corollary 1 we obtain
%the corresponding intervals of $x$ for which $\rho$ is genuine tripartite entangled. The results
%are shown in Table \ref{tab:1}. Explicitly,
%which shows that $\rho$ is genuine tripartite entangled for
Let $f_1(x)=T(\rho)-J_1=5x-(2+\sqrt{3})$, using Corollary 1 we have $\rho$ is genuine entangled if $f_1(x)>0$, i.e. $0.7464<x\leq1$. %Set $g_1(x)=\frac{1}{12}(\sqrt{66}x-6)$, Theorem 2 in \cite{lmj} says that when $g_1(x)>0$, $\rho$ is genuine entangled, i.e. for $0.7385<x\leq1$.
Theorem 2 in \cite{dgh} implies that $\rho$ is genuine entangled if $g_1(x)=3.26x-\frac{6+\sqrt{3}}{3}>0$, i.e. $0.791<x\leq1$. Our corollary can detect more genuine entangled, see the comparison in Fig. \ref{fig:1}.
%\begin{table}[!htb]
%\caption{$T(\rho_W)$, $K_1$ and the range of GME of the state (\ref{14}) for different $\alpha$ and $\beta$.}
%\label{tab:1}
%\centering
%\begin{tabular}{cccc}
%%\toprule
%\hline\noalign{\smallskip}
%\ & $T(\rho_W)$ & $K_1$ & the range of GME \\
%%\midrule
%$\alpha=1,\beta=1$ & $3.7177x$ & $1+\sqrt{3}$ & $0.7349<x\leq1$ \\
%$\alpha=\frac{1}{2},\beta=2$ & $6.5825x$ & $\frac{1}{2}+2\sqrt{3}$ & $0.6022<x\leq1$\\
%$\alpha=\frac{1}{10},\beta=2$ & $6.5225x$ & $\frac{1}{10}+2\sqrt{3}$ & $0.5464<x\leq1$ \\
%%\bottomrule
%\end{tabular}
%\end{table}

%\textit{\textbf{Example 2}} Consider the quantum state $\rho\in H_1^3\otimes H_2^3\otimes H_3^2$,
%\begin{equation}\label{15}
%\rho=\frac{1-x}{18}I_{18}+x|\varphi\rangle\langle\varphi|,
%\end{equation}
%where $|\varphi\rangle=\frac{1}{\sqrt{5}}[(|10\rangle+|21\rangle)|0\rangle
%+(|00\rangle+|11\rangle+|22\rangle)|1\rangle]$, $0\leq x\leq1$, $I_{18}$ is the $18\times18$ identity matrix.
%By Theorem 1, we can determine the range of $x$ where $\rho$ is surely entangled. Table \ref{tab:2} shows that when $\alpha=0$, $\beta=1$,
%our criterion detects the entanglement for $0.3405<x\leq1$, which is better than the result $0.35\leq x\leq1$ given in \cite{lc}.
%Set $f_2(x)=T(\rho)-\frac{1}{3}(2\sqrt{232}+\sqrt{192})=15.5320x-\frac{1}{3}(2\sqrt{232}+\sqrt{192})$, therefore $\rho$ is genuine tripartite entangled when $f_2(x)>0$, i.e. $0.9511<x\leq1$. Note that the criterion given in \cite{lc} cannot detect genuine entanglement.
%\begin{table}[!htb]
%\caption{The entanglement regions of $\rho$ as given by Theorem 1.}
%\label{tab:2}
%\centering
%\begin{tabular}{ccc}
%\hline\noalign{\smallskip}
%\  & $\|N^{2|13}\|_{tr}$ & The range of entanglement \\
%$\alpha=1, \beta=1$ & $10.5292x$ & $0.4852<x\leq1$ \\
%$\alpha=\frac{1}{2}, \beta=2$ & $18.4650x$ & $0.3909<x\leq1$\\
%$\alpha=0, \beta=1$ & $9.1321x$ & $0.3405<x\leq1$\\
%\end{tabular}
%\end{table}

\section{Entanglement for multipartite quantum state}
Now we consider entanglement of $n$-partite quantum systems. Let $\{A_{u_s}^{(s)}\}$ $(u_s=0,\cdots,d_s^2-1)$ be the generalized Pauli operators of the $s$th $d_s$-dimensional Hilbert space $H_s^{d_s}$. Any quantum state
$\rho\in H_1^{d_1}\otimes H_2^{d_2}\otimes\cdots\otimes H_n^{d_n}$ has the generalized Pauli operators representation:
\begin{equation}\label{16}
\rho=\frac{1}{d_1d_2\cdots d_n}\sum\limits_{s=1}^n\sum_{u_s=0}^{d_s^2-1}t_{u_1,u_2,\cdots,u_n}A_{u_1}^{(1)}\otimes A_{u_2}^{(2)}\otimes \cdots\otimes A_{u_n}^{(n)}
\end{equation}
where $A_0^{(s)}=I_{d_s}$$(s=1,\cdots,n)$, $t_{u_1,u_2,\cdots,u_n}=tr(\rho (A_{u_1}^{(1)})^{\dagger}\otimes(A_{u_2}^{(2)})^{\dagger}\otimes \cdots\otimes (A_{u_n}^{(n)})^{\dagger})$ are complex coefficients.
Let $T^{(l_1\cdots l_k)}$ be the vectors with entries $t_{u_{l_1},\cdots,u_{l_k},\cdots,0}$, $u_{l_1},\cdots,u_{l_k}\neq0$ and $1\leq l_1<\cdots< l_k\leq n$.
We have
%\begin{equation*}
%\begin{split}
$$\|T^{(1)}\|^2=\sum_{u_1=1}^{d_1^2-1}t_{u_1,\cdots,0}t_{u_1,\cdots,0}^{*},$$
$$\cdots,$$
$$\|T^{(l_1\cdots l_k)}\|^2=\sum\limits_{s=1}^{k}\sum_{u_{l_s}=1}^{d_{l_s}^2-1}t_{u_{l_1}\cdots u_{l_k}\cdots0}t_{u_{l_1}\cdots u_{l_k}\cdots0}^{*},$$
$$\cdots,$$
$$\|T^{(12\cdots n)}\|^2=\sum\limits_{s=1}^n\sum_{u_s=1}^{d_s^2-1}t_{u_1\cdots u_n}t_{u_1\cdots u_n}^{*},$$
%\end{split}
%\end{equation*}
where $*$ represents the conjugate.
Set
$$A_1=\|T^{(1)}\|^2+\cdots+\|T^{(n)}\|^2,$$
$$A_2=\|T^{(12)}\|^2+\cdots+\|T^{((n-1),n)}\|^2,$$ $$\cdots,$$ $$A_n=\|T^{(1\cdots n)}\|^2.$$

\begin{lemma}\label{3}
Let $\rho\in H_1^{d_1}\otimes H_2^{d_2}\otimes\cdots\otimes H_n^{d_n}$ $(n\geq2)$ be a $n$-partite pure quantum state. Then
\begin{equation}\label{17}
\|T^{(12\cdots n)}\|^2\leq
\frac{d_1\cdots d_n(n-1-\sum\limits_{s=1}^n\frac{1}{d_s^2})+1}{n-1}.
\end{equation}
\end{lemma}
{\it Proof}~~It's enough to show the lemma for a pure state $\rho$, where we have $tr(\rho^2)=1$ and $tr(\rho_{l_1}^2)=tr(\rho_{l_2\cdots l_n}^2)$ for any distinct indices $l_1, \ldots, l_n\in\{1,2,\cdots,n\}$. Here
$\rho_{l_1}$ and $\rho_{l_2\cdots l_n}$ are the reduced states for the subsystem $H_{l_1}^{d_{l_1}}$ and $H_{l_2}^{d_{l_2}}\otimes\cdots\otimes H_{l_n}^{d_{l_n}}$. Therefore, we have
\begin{equation}\label{18}
tr(\rho^2)=\frac{1}{d_1d_2\cdots d_n}(1+A_1+\cdots+A_n)=1,
\end{equation}
and
\begin{equation}\label{19}
\frac{1}{d_{l_1}}(1+\|T^{(l_1)}\|^2)=\frac{1}{d_{l_2}\cdots d_{l_n}}(1+\|T^{(l_2)}\|^2+\cdots+\|T^{(l_n)}\|^2+\cdots+\|T^{(l_2\cdots l_n)}\|^2).
\end{equation}
Since $\sum\limits_{l_1=1}^n\frac{1}{d_{l_1}}tr(\rho_{l_1}^2)=\sum\limits_{l_1=1}^n\frac{1}{d_{l_1}}tr(\rho_{l_2\cdots l_n}^2)$,
we get that
\begin{equation*}
\sum\limits_{l_1=1}^n\frac{1}{d_{l_1}^2}(1+\|T^{(l_1)}\|^2)=\frac{1}{d_{1}\cdots d_{n}}[n+(n-1)A_1+(n-2)A_2+\cdots+A_{n-1}].
\end{equation*}
Therefore,
\begin{equation}\label{20}
A_1=\frac{d_{1}\cdots d_{n}}{n-1}\sum\limits_{s=1}^n\frac{1}{d_{s}^2}(1+\|T^{(s)}\|^2)
-\frac{n}{n-1}-\frac{n-2}{n-1}A_2-\frac{n-3}{n-1}A_3-\cdots-\frac{1}{n-1}A_{n-1}.
\end{equation}
Substituting (\ref{20}) into (\ref{18}), we get
\begin{equation}\label{21}
\begin{split}
A_n=&d_1\cdots d_n-1-\frac{1}{n-1}\left(d_{1}\cdots d_{n}\sum\limits_{s=1}^n\frac{1}{d_{s}^2}(1+\|T^{(s)}\|^2)
-n\right)-\frac{1}{n-1}A_2\\
&-\frac{2}{n-1}A_3-\cdots\frac{n-2}{n-1}A_{n-1}\\
\leq&\frac{d_1\cdots d_n(n-1-\sum\limits_{s=1}^n\frac{1}{d_s^2})+1}{n-1}
\end{split}
\end{equation}
\qed

Let $\rho$ be a $n$-partite state $\rho\in H_1^{d_1}\otimes H_2^{d_2}\otimes\cdots\otimes H_n^{d_n}$ represented as \eqref{16},
%suppose $\rho$ is separable under the bipartition $l_1\cdots l_{k-1}|l_k\cdots l_{n}$.
for real number $\alpha$, $\beta$ and distinct
indices $l_1, \ldots, l_n\in\{1, 2, \cdots, n\}$,
%$l_1<\cdots<l_{k-1}, l_k<\cdots <l_{n}\in\{1, 2, \cdots, n\}$,
set
\begin{equation}\label{22}
N^{l_1\cdots l_{k-1}|l_k\cdots l_{n}}=\alpha S_0^{l_1\cdots \l_{k-1}|l_k}+\beta S^{l_1\cdots l_{k-1}|l_k\cdots l_n},
\end{equation}
for $k-1=1, 2,\cdots, [n/2]$, the smallest integer less or equal to $n/2$. Let $T^{(l_1\cdots l_k)}$ be the $(d^2_{l_1}-1)\cdots (d^2_{l_k}-1)$-dimensional column vector with entries $t_{u_{l_1}\cdots u_{l_k}\cdots0}$
associated with the generalized Pauli operators representation of $\rho$, and define
%and $m=\frac{n}{2}$ when $n$ is even, and $m=\frac{n-1}{2}$ when $n$ is odd.
$S_0^{l_1\cdots l_{k-1}|l_k}$ to be the block matrix
$S_0^{l_1\cdots \l_{k-1}|\l_k}=[S^{l_1\cdots \l_{k-1}|\l_k}~~O_{l_1\cdots l_{k-1}}]$, where
$S^{l_1\cdots l_{k-1} l_k}=
%T^{(l_1\cdots l_{k-1})}(T^{(l_k)})^{\dagger}=
([t_{u_{l_1}\cdots u_{l_k}\cdots0}])$ is the
$\prod\limits_{s=1}^{k-1}(d_{l_s}^2-1)\times (d_{l_k}^2-1)$ matrix and
$O_{l_1\cdots l_{k-1}}$ is the $\prod\limits_{s=1}^{k-1}(d_{l_s}^2-1)\times
[\prod\limits_{s=k}^{n}(d_{l_s}^2-1)-(d_{l_k}^2-1)]$
zero matrix, and $S^{l_1\cdots l_{k-1}|l_k\cdots l_n}
%=T^{(l_1\cdots l_{k-1})}(T^{(l_k\cdots l_n)})^{\dagger}
=[t_{u_1,\cdots, u_n}]$ to be a $\prod\limits_{s=1}^{k-1}(d_{l_s}^2-1)\times
\prod\limits_{s=k}^{n}(d_{l_s}^2-1)$ matrix. For example, when $\rho\in H_1^{2}\otimes H_2^{2}\otimes H_3^{2}\otimes H_4^{3}$, $N^{13|24}=\alpha S_0^{13|2}+\beta S^{13|24}$, where
$$S^{13|2}=\left[
           \begin{array}{ccc}
             t_{1,1,1,0}~~ & t_{1,2,1,0}~~ & t_{1,3,1,0} \\
             t_{1,1,2,0}~~ & t_{1,2,2,0}~~ & t_{1,3,2,0} \\
             t_{1,1,3,0}~~ & t_{1,2,3,0}~~ & t_{1,3,3,0} \\
             \vdots & \vdots & \vdots \\
             t_{3,1,3,0}~~ & t_{3,2,3,0}~~ & t_{3,3,3,0} \\
           \end{array}
         \right],~~
S^{13|24}=\left[
           \begin{array}{cccccc}
             t_{1,1,1,1}~ & t_{1,1,1,2} & \cdots & t_{1,1,1,8} & \cdots & t_{1,3,1,8}\\
             t_{1,1,2,1}~ & t_{1,1,2,2} & \cdots & t_{1,1,2,8} & \cdots & t_{1,3,2,8}\\
             t_{1,1,3,1}~ & t_{1,1,3,2} & \cdots  & \cdot  & \cdots  & \cdot \\
             \vdots~ & \vdots & \vdots  & \vdots  & \vdots  & \vdots  \\
             t_{3,1,3,1}~ & t_{3,1,3,2} & \cdots  & \cdot  & \cdots  & \cdot  \\
           \end{array}
         \right].
$$

\begin{theorem}\label{3} Fix $\alpha, \beta$ as above.
If the $n$-partite state $\rho\in H_1^{d_1}\otimes H_2^{d_2}\otimes\cdots\otimes H_n^{d_n}$ is separable under the bipartition $l_1\cdots l_{k-1}|l_k\cdots l_{n}$, then we have that\\
(i) $\|N^{l_1|l_{2}\cdots l_{n}}\|_{tr}\leq M_{l_1}$;\\
(ii) $\|N^{l_1\cdots l_{k-1}|l_k\cdots l_{n}}\|_{tr}\leq M_{l_1\cdots l_{k-1}}$ $(k\geq3)$;\\
where
\begin{footnotesize}
\begin{equation*}
M_{l_1}=\sqrt{d_{l_1}-1}\left(|\alpha|\sqrt{d_{l_2}-1}+|\beta|
\sqrt{\frac{d_{l_2}\cdots d_{l_n}(n-2-\sum\limits_{s=2}^nd_{l_s}^{-2})+1}{n-2}}\right),
\end{equation*}
\begin{equation*}
M_{l_1\cdots l_{k-1}}=\sqrt{\frac{d_{l_1}\cdots d_{l_{k-1}}(k-2-\sum\limits_{s=1}^{k-1}d_{l_s}^{-2})+1}{k-2}}
\left(|\alpha|\sqrt{d_{l_k}-1}+|\beta|\sqrt{\frac{d_{l_k}\cdots d_{l_n}(n-k-\sum\limits_{s=k}^nd_{l_s}^{-2})+1}{n-k}}\right).
\end{equation*}
\end{footnotesize}
\end{theorem}
{\it Proof}~~$(i)$ If the $n$-partite mixed state is separable under the bipartition $l_1|l_2\cdots l_{n}$, it can be expressed as
\begin{equation}\label{23}
\rho_{l_1|l_2\cdots l_{n}}=\sum\limits_s p_s\rho_{l_1}^s\otimes\rho_{l_2\cdots l_{n}}^s, \ 0<p_s\leq1, \sum\limits_s p_s=1,
\end{equation}
where
\begin{equation}\label{24}
\rho_{l_1}^s=\frac{1}{d_{l_1}}\sum_{u_{l_1}=0}^{d_{l_1}^2-1} t_{u_{l_1}}^sA_{u_{l_1}}^{(l_1)},
\end{equation}
\begin{equation}\label{25}
\rho_{l_2\cdots l_{n}}^s=\frac{1}{d_{l_2}\cdots d_{l_{n}}}\sum\limits_{q=2}^{n}\sum_{u_{l_q}=0}^{d_{l_q}^2-1}t_{u_{l_2},\cdots,u_{l_n}}^s
A_{u_{l_2}}^{(l_2)}\otimes \cdots\otimes A_{u_{l_{n}}}^{(l_{n})}
\end{equation}
Then,
\begin{equation}\label{26}
S^{l_1|l_2}=\sum\limits_sp_sT_s^{(l_1)}(T_s^{(l_2)})^{t},~~
S^{l_1|l_2\cdots l_n}=\sum\limits_sp_sT_s^{(l_1)}(T_s^{(l_2\cdots l_n)})^{t}.
\end{equation}
By Lemma 1 and Lemma 3, we have
\begin{equation*}
\begin{split}
\|N^{l_1|l_2\cdots l_{n}}\|_{tr}&\leq\sum_sp_s(|\alpha|\|T_s^{(l_1)}\|\|T_s^{(l_2)}\|+|\beta|\|T_s^{(l_1)}\|\|T_s^{(l_2\cdots l_n)}\|)\\
&\leq\sqrt{d_{l_1}-1}\left(|\alpha|\sqrt{d_{l_2}-1}+|\beta|
\sqrt{\frac{d_{l_2}\cdots d_{l_n}(n-2-\sum\limits_{s=2}^n\frac{1}{d_{l_s}^2})+1}{n-2}}\right)\\
&=M_{l_1},
\end{split}
\end{equation*}
where we have used $\|A+B\|_{tr}\leq\|A\|_{tr}+\|B\|_{tr}$ for matrices $A$ and $B$ and $\||a\rangle\langle b|\|_{tr}=\||a\rangle\|\||b\rangle\|$
for vectors $|a\rangle$ and $|b\rangle$.\\
$(ii)$ If $\rho$ is separable under the bipartition $l_1\cdots l_{k-1}|l_k\cdots l_{n}$, it can be expressed as
\begin{equation}\label{27}
\rho_{l_1\cdots l_{k-1}|l_k\cdots l_{n}}=\sum\limits_s p_s\rho_{l_1\cdots l_{k-1}}^s\otimes\rho_{l_k\cdots l_{n}}^s, 0<p_s\leq1, \sum\limits_s p_s=1,
\end{equation}
where
\begin{equation}\label{28}
\rho_{l_1\cdots l_{k-1}}^s=\frac{1}{d_{l_1}\cdots d_{l_{k-1}}}\sum\limits_{p=1}^{k-1}\sum_{u_{l_p}=0}^{d_{l_p}^2-1}
t_{u_{l_1},\cdots,u_{l_{k-1}}}^s
A_{u_{l_1}}^{(l_1)}\otimes \cdots\otimes A_{u_{l_{k-1}}}^{(l_{k-1})},
\end{equation}
\begin{equation}\label{29}
\rho_{l_k\cdots l_{n}}^s=\frac{1}{d_{l_k}\cdots d_{l_{n}}}\sum\limits_{q=k}^{n}\sum_{u_{l_q}=0}^{d_{l_q}^2-1}t_{u_{l_k},\cdots,u_{l_{n}}}^s
A_{u_{l_k}}^{(l_k)}\otimes \cdots\otimes A_{u_{l_{n}}}^{(l_{n})}.
\end{equation}
Then,
\begin{equation}\label{30}
S^{l_1\cdots l_{k-1}|l_k}=\sum\limits_sp_sT_s^{(l_1\cdots \l_{k-1})}(T_s^{(l_k)})^{t},~~
S^{l_1\cdots l_{k-1}|l_k\cdots l_n}=\sum\limits_sp_sT_s^{(l_1\cdots \l_{k-1})}(T_s^{(l_k\cdots l_n)})^{t}.
\end{equation}
Similarly, we get
\begin{equation*}
\begin{split}
&\|N^{l_1\cdots l_{k-1}|l_k\cdots l_{n}}\|_{tr}\\
\leq&\sum_sp_s(|\alpha|\|T_s^{(l_1\cdots l_{k-1})}\|\|T_s^{(l_k)}\|+|\beta|\|T_s^{(l_1\cdots l_{k-1})}\|\|T_s^{(l_k\cdots l_n)}\|)\\
\leq&\sqrt{\frac{d_{l_1}\cdots d_{l_{k-1}}(k-2-\sum\limits_{s=1}^{k-1}\frac{1}{d_{l_s}^2})+1}{k-2}}
[|\alpha|\sqrt{d_{l_k}-1}+|\beta|\sqrt{\frac{d_{l_k}\cdots d_{l_{n}}(n-k-\sum\limits_{s=k}^n\frac{1}{d_{l_s}^2})+1}{n-k}}]\\
=&M_{l_1\cdots l_{k-1}}.
\end{split}
\end{equation*}
\qed

\textit{\textbf{Example 2}} Consider the quantum state $\rho\in H_1^3\otimes H_2^3\otimes H_3^2$,
\begin{equation}\label{15}
\rho=\frac{1-x}{18}I_{18}+x|\varphi\rangle\langle\varphi|,
\end{equation}
where $|\varphi\rangle=\frac{1}{\sqrt{5}}[(|10\rangle+|21\rangle)|0\rangle
+(|00\rangle+|11\rangle+|22\rangle)|1\rangle]$, $0\leq x\leq1$, $I_{18}$ is the $18\times18$ identity
matrix. By Theorem 3 $(i)$, we can determine the range of $x$ where $\rho$ is surely entangled. Table\ref{tab:2} shows that when $\alpha=0$, $\beta=1$,
our criterion detects the entanglement for $0.3405<x\leq1$, which is better than the result $0.35\leq
x\leq1$given in \cite{lc}.

\begin{table}[!htb]
\caption{The entanglement regions of $\rho$ as given by Theorem 3.}
\label{tab:2}
\centering
\begin{tabular}{ccc}
\hline\noalign{\smallskip}
\  & $\|N^{2|13}\|_{tr}$ & The range of entanglement \\
\noalign{\smallskip}\hline\noalign{\smallskip}
$\alpha=1, \beta=1$ & $10.5292x$ & $0.4852<x\leq1$ \\
$\alpha=\frac{1}{2}, \beta=2$ & $18.4650x$ & $0.3909<x\leq1$\\
$\alpha=0, \beta=1$ & $9.1321x$ & $0.3405<x\leq1$\\
\noalign{\smallskip}\hline
\end{tabular}
\end{table}

\begin{figure}[!htb]
  \centering
  \includegraphics[width=0.75\textwidth]{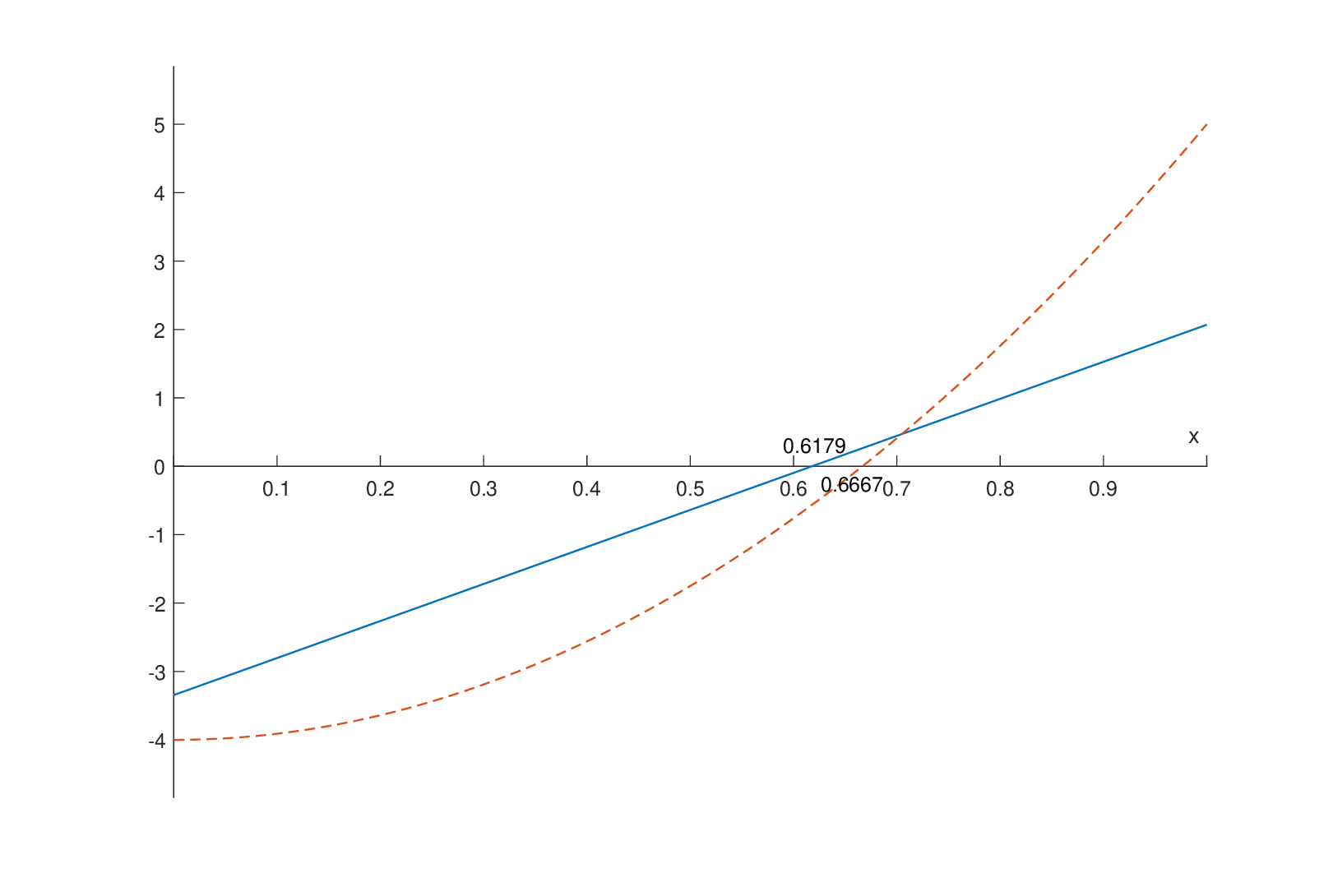}
  \caption{$f_2(x)$ from our result (solid straight line), $g_2(x)$ from Theorem 3 in \cite{lww} (dashed curve line). }
  \label{fig:2}
\end{figure}

\textit{\textbf{Example 3}} Consider the four-qubit state $\rho\in H_1^2\otimes H_2^2\otimes H_3^2\otimes H_4^2$,
\begin{equation}\label{32}
 \rho=x|\psi\rangle\langle\psi|+\frac{1-x}{16}I_{16},
\end{equation}
where $|\psi\rangle=\frac{1}{\sqrt{2}}(|0000\rangle+|1111\rangle)$, $0\leq x\leq1$,
$I_{16}$ is the $16\times16$ identity matrix.
Using Theorem 3 $(i)$ with $\alpha=1, \beta=1$, we set
$f_2(x)=\|N^{l_1|l_2l_3l_4}\|_{tr}-(1+\sqrt{\frac{11}{2}})=(4+\sqrt{2})x-(1+\sqrt{\frac{11}{2}})$, $\rho$ is not separable under the bipartition $l_1|l_2l_3l_4$ for $f_2(x)>0$, i.e. $0.6179<x\leq1$, while according to Theorem 3 in \cite{lww}, $\rho$ is not separable under the bipartition $l_1|l_2l_3l_4$ for $g_2(x)=9x^2-4>0$, i.e. $0.6667<x\leq1$. Fig. \ref{fig:2} shows that our method detects more entanglement. %$f_4(x)=\|N^{l_1l_2|l_3l_4}\|_{tr}-3=5x-(3+\sqrt{3})$ for the bipartition $l_1l_2|l_3l_4$, $\rho$ is not separable under the bipartition $l_1l_2|l_3l_4$ for $f_4(x)>0$, i.e. $0.9464<x\leq1$.

%Now using Corollary 2 and set $\alpha=1, \beta=1$, we get that %$f_4(x)=T(\rho)-J_2=\frac{23+2\sqrt{2}}{5}x-\frac{11+\sqrt{22}+3\sqrt{3}}{5}$,
%$\rho$ is genuine entanglement for $ f_4(x)>0$, i.e.$0.8087<x\leq1$, while the criterion given in %\cite{lww} cannot detect the genuine four-qubits entanglement at all.

\section{Conclusions}
By adopting the representation with generalized Pauli operators of density matrices, we have come up with several general tests to judge genuine entanglement for tripartite quantum systems. Our approach starts with some finer upper bounds for the norms of correlation tensors by using the generalized Pauli operators presentation, then we have obtained the entanglement criteria for genuine tripartite quantum states based on certain matrices constructed by the correlation tensor of the density matrices. We also conducted conclusion to detect entanglement in arbitrary dimensional multipartite quantum states. Compared with previously available criteria, ours can detect more situations, and these are explained in details with several examples.
%Detailed examples have been shown that our criteria can detect more genuine entangled states.

\textbf{Acknowledgments} This work is supported by the National Natural Science Foundation of China under grant nos. 12075159, 12126351 and 12171044, Simons Foundation under grant no. 523868, Beijing Natural Science Foundation (grant no. Z190005), Academy for Multidisciplinary Studies, Capital Normal University, and Shenzhen Institute for Quantum Science and Engineering, Southern University of Science and Technology (no. SIQSE202001), and the Academician Innovation Platform of Hainan Province.

\noindent\textbf {Data Availability Statements} All data generated or analysed during this study are available from the corresponding author on reasonable request.

\end{document}